\documentclass[12pt]{article}
\usepackage{amsfonts}
\usepackage{amsmath}
\usepackage{amssymb}
\usepackage{graphicx}
\usepackage{color}
\usepackage[all, knot]{xy}
\usepackage{tikz}

\usepackage{ulem}

\usepackage[utf8]{inputenc}
\usepackage{epstopdf}
\usepackage[footnotesize]{caption}
\usepackage{amsthm}
\usepackage{enumitem}
\usepackage{mathrsfs}

\usepackage[margin=3cm]{geometry}

\def \be {\begin{equation}}
\def \ee {\end{equation}}
\def \bea {\begin{eqnarray}}
\def \eea {\end{eqnarray}}
\def \nn {\nonumber}

\def \rr {\raise.35ex\hbox{\small $\prime$}\kern-.17em{\mbox{\large $\imath$}}}

\def \dels {\partial\kern-.6em /\kern.1em}
\def \As {{A\kern-.5em / \kern.5em}}
\def \Ds {D\kern-.7em / \kern.5em}

\def \ks {k\kern-.5em /}
\def \ls {l\kern-.5em /}

\newcommand{\ci}[1]{}

\newcommand{\lb}{\left(}
\newcommand{\rb}{\right)}

\newcommand{\lsb}{\left[}
\newcommand{\rsb}{\right]}

\newcommand{\ba}{\begin{eqnarray}}
\newcommand{\ea}{\end{eqnarray}}
\newcommand{\bal}{\begin{align}}
\newcommand{\eal}{\end{align}}
\newcommand{\bay}[1]{\left(\begin{array}{#1}}
\newcommand{\eay}{\end{array}\right)}

\newcommand{\ie}{\textit{i.e.}, }

\newcommand{\zt}[1]{\textrm{#1}}

\def\rmd{{\rm d}}

\def\xd{{\delta}}
\def\xD{{\Delta}}
\def\xe{{\epsilon}}

\def\xS{{\Sigma}}

\def\CO{{\cal O}}

\def\CV{{\cal V}}

\newcommand{\hide}[1]{}

\newlist{axioms}{enumerate}{2}
\setlist[axioms,1]{label=\textbf{A\arabic{axiomsi}.}, ref=A\arabic{axiomsi}}
\setlist[axioms,2]{label=\textbf{A\arabic{axiomsi}\rlap{\myEnumCounter{axiomsii}}.},%
                   ref=A\arabic{axiomsi}\myEnumCounter{axiomsii},%
                   align=parleft,%
                   leftmargin=0em,%
                   itemsep=1.4ex,%
                   before={\stepcounter{axiomsi}}}

  \usetikzlibrary{decorations.markings}

\begin{document}

\begin{titlepage}
\begin{center}

\textbf{\LARGE
The Probe of Curvature in\\
 the Lorentzian AdS$_2$/CFT$_1$ Correspondence
\vskip.3cm
}
\vskip .5in
{\large
Xing Huang$^{a,b}$ \footnote{e-mail address: xingavatar@gmail.com} and
Chen-Te Ma$^{c,d,e}$ \footnote{e-mail address: yefgst@gmail.com}
\\
\vskip 1mm
}
{\sl
$^a$
Institute of Modern Physics, Northwest University, Xi'an 710069, China.
\\
$^b$
Shaanxi Key Laboratory for Theoretical Physics Frontiers, Xi'an 710069, China.
\\
$^c$
Institute of Quantum Matter,\\
 School of Physics and Telecommunication Engineering,\\
 South China Normal University, Guangzhou 510006, Guangdong, China.
\\
$^d$
The Laboratory for Quantum Gravity and Strings,\\
 Department of Mathematics and Applied Mathematics,\\
University of Cape Town, Private Bag, Rondebosch 7700, South Africa.
\\
$^e$
Department of Physics and Center for Theoretical Sciences, \\
National Taiwan University, Taipei 10617, Taiwan, R.O.C..
}\\
\vskip 1mm
\vspace{40pt}
\end{center}

\begin{abstract}
We establish the Lorentzian AdS$_2$/CFT$_1$ correspondence from a reconstruction of all bulk points through the kinematic-space approach. The OPE block is exactly a bulk local operator. We formulate the correspondence between the bulk propagator in the non-interacting scalar field theory and the conformal block in CFT$_1$. When we consider the stress tensor, the variation probes the variation of AdS$_2$ metric. The reparameterization provides the asymptotic boundary of the bulk spacetime as in the derivation of the Schwarzian theory from two-dimensional dilaton gravity theory. Finally, we find the AdS$_2$ Riemann curvature tensor based on the above consistent check. 
\end{abstract}
\end{titlepage}

\section{Introduction}
\label{sec:1}
\noindent
The holographic correspondence states that the physical degrees of freedom in quantum gravity theory has an equivalent description from the boundary. Nowadays, the only perturbatively computable quantum gravity theory without any known inconsistency is string theory. String theory lives on two-dimensional worldsheet, and the fluctuation of target space without conformal anomaly provides a theory containing Einstein gravity theory \cite{Strominger:2017zoo}. Based on the implication of string theory, the four-dimensional conformal field theory (CFT$_4$), ${\cal N}=4$ supersymmetric Yang-Mills theory is dual to string theory on the five-dimensional anti-de Sitter spacetime times a five-dimensional sphere manifold (AdS$_5\times S_5$) \cite{Maldacena:1997re}. This motivates the more generic and testable holographic-conjecture, AdS/CFT correspondence. 
\\

\noindent
The quantum fluctuation in Einstein gravity theory has the known issue in the renormalizability. Therefore, the ultraviolet (UV) information cannot be probed so far. Through the AdS/CFT correspondence, the AdS quantum gravity theory can be defined by CFT. Since CFT is scale-invariant, the UV divergence is not problematic. Therefore, AdS/CFT correspondence or CFT becomes an important direction in the study of quantum gravity theory. Because the techniques of CFT are widely used in condensed matter physics, the correspondence is also useful in the application of strongly correlated condensed matter system \cite{Ma:2018efs}.  
\\

To probe the bulk physics from the boundary theory, the unavoidable study is the operator dictionary in the AdS/CFT correspondence. To reach the goal, we should first consider the pairs of operator insertions or bi-local operators. We can use the operator product expansion (OPE) to organize the CFT bi-local operators. Through the geodesic Witten diagrams, people obtained the conformal block for the spacelike-separated points \cite{Hijano:2015zsa}.  For the time-like separated points, the OPE block corresponds to the codimension-two surface operator \cite{Czech:2016xec, deBoer:2016pqk}, which is localized on the bulk codimension-two minimum surface on a given time slice that serves as the boundary of casual wedge specified by two points. Each conformal block for spacelike-separated points is equivalent to the correlator of two OPE-blocks \cite{Czech:2016xec, deBoer:2016pqk}. The conformal kinematics of a scalar OPE block can be written as the Klein-Gordon equation in the kinematic space. Here we continue to use the term kinematic space to refer to the spacetime specified by a pair of points.
The OPE block of the stress tensor also exactly corresponds to the modular Hamiltonian \cite{Casini:2011kv} for the natural study of holographic entanglement entropy, which obeys the Liouville and Toda equations in the gravity theory and its spin-3 extension \cite{deBoer:2016pqk}.
\\

When one considers the Lorentzian CFT$_1$, there is no longer any codimension-two extended surface on a given time slice whose the area has the interpretation of entanglement entropy. Therefore, we need to renew the holographic study. Since the Lorentzian CFT$_1$ only has a time direction, we directly use two boundary points to reconstruct each bulk point. \footnote{We noticed that the same two-to-one mapping was also used in \cite{Blommaert:2019hjr}. Nevertheless, our studies of the bulk, based on the OPE blocks, are very different. It was also used in the construction of boundary kinematic space \cite{Callebaut:2018xfu}.} In other words, each point in the kinematic space is a bulk point. This makes the kinematic space very interesting as it is, in fact, the AdS spacetime itself. This also leads to the new conclusion that the OPE block is the bulk local operator for the convenient holographic correspondence, which will be supported by our findings on the conformal block \cite{Dolan:2003hv, SimmonsDuffin:2012uy} and the modular Hamiltonian.
We also use the reparametrization to obtain the asymptotic boundary as in the derivation of the Schwarzian theory \cite{Jensen:2016pah}.
The Schwarzian theory is the classical effective action of the two-dimensional dilaton gravity theory.
This has the holographic evidence from the consistent dilaton solution given by the bulk and boundary sides \cite{Maldacena:2016upp}.
 Since the OPE block cannot detect the interacting information of metric, we also study the vacuum Virasoro OPE block. The Hawking temperature in the AdS$_3$ \cite{Fitzpatrick:2015foa} and the bulk operator \cite{daCunha:2016crm, Guica:2016pid, Fitzpatrick:2016mtp} were also determined from the vacuum Virasoro OPE block. Therefore, this precisely determines the relation between the AdS$_2$ metric and the modular Hamiltonian.
\\

Recently, modular Berry connection \cite{Czech:2017zfq} is proposed as a natural definition to the connection of the modular Berry transport. This provides the connection of the Riemann curvature tensor in the kinematic space from CFT$_2$ \cite{Czech:2019vih}. The central question that we would like to address in this letter is the following: {\it How do we probe the AdS$_2$ Riemann curvature tensor from the Lorentzian CFT$_1$?}
From the consistent construction of the Lorentzian AdS$_2$/CFT$_1$ correspondence, we obtain the AdS$_2$ Riemann curvature tensor. We would like to emphasize that some properties of kinematic spaces and OPE blocks, particularly what we can learn from them about the spacetime, are very distinct in different dimensions, which becomes apparent in the study of modular Hamiltonian. The current example offers the best-case scenario as everything follows from the kinematic space, which is the reason why this special case is worth all the extensive study. 

\section{OPE Block and Conformal Block}
\label{sec:2}
The OPE block $B^{jk}_l (x_1, x_2)$ \cite{Czech:2016xec} follows from the operator product expansion (OPE) of the operators ${\cal O}_j(x_1)$ and ${\cal O}_k(x_2)$
\bea
{\cal O}_j(x_1){\cal O}_k(x_2)=\sum_l c_{jkl}\big(x_1-x_2, \partial\big){\cal O}_l(x_2),
\eea
in which $c_{jkl}$ takes into account the contribution from the descendants
of operator ${\cal O}_k(x_1, x_2)$. The OPE block is defined as the contribution
from a particular channel of primary operator to the OPE of the operators ${\cal O}_j(x_1)$ and ${\cal O}_k(x_2)$
\bea
\label{OPEblockdef} {\cal O}_j(x_1){\cal O}_k(x_2)\equiv |x_1-x_2|^{-\Delta_j-\Delta_k}\sum_l C_{jkl}B_l^{jk}(x_1, x_2),
\nn\\
\eea
where $\Delta_{j}$ and $\Delta_k$ are conformal dimensions of the operators, ${\cal O}_{j}$ and ${\cal O}_{k}$, and $C_{jkl}$ are the OPE coefficients.
\\

Here we use the codimension-two surface in the time direction. Therefore, the OPE block should correspond to a bulk local operator in this holographic set-up. The Lorentzian AdS$_2$ metric is $d s_{2l}^2 = (-d t^2  + d z^2)/z^2$. The light cone of a boundary point becomes a single light ray in the bulk. Therefore, the past light ray of the boundary point $\tau_2$ and the future light ray of the boundary point $\tau_1$ (assuming $\tau_2>\tau_1$) meet at the following bulk point in the Lorentzian AdS$_2$ metric. In general, the bulk point is determined by \cite{Blommaert:2019hjr}:
\bea
\label{bulkcoord}
t =\frac{1}{2} (\tau_1 + \tau_2),\qquad z= \frac{1}{2} |\tau_1 - \tau_2|.
\eea
In summary, the OPE block or the codimension-two surface operator from two boundary operators at $\tau_1$ and $\tau_2$ becomes a bulk local operator, whose position is uniquely determined by \eqref{bulkcoord}. To see this, we will
establish equality between the Lorentzian AdS$_2$ bulk propagator and the conformal block in the Lorentzian CFT$_1$. We will begin from the known Euclidean results and perform the Wick rotation to the Lorentzian case. Ones can also directly solve the wave equations in the Lorentzian spacetime without the Wick rotation.
\\

The action of the Euclidean AdS$_2$ bulk theory is given by
\bea
S_{\mathrm{scalar}}=\int d^2x\sqrt{|\det g_{\mu\nu}|}\ \bigg(\frac{1}{2}g^{\rho\sigma}\partial_{\rho}\phi\partial_{\sigma}\phi+\frac{m^2}{2}\phi^2\bigg),
\nn\\
\eea
where $m^2\equiv-\Lambda\Delta(\Delta-1)$ is the mass square of the scalar field $\phi$, and $\Delta$ is the conformal dimension of a boundary operator. The AdS$_2$ spacetime indices are labeled by $\rho$ and $\sigma$. The Euclidean metric field is defined by:
 \bea
 ds_{2dw}^2&\equiv& 2e^{2\rho_E}dx^+dx^-\equiv -dX_1^2-dX_2^2+dX_3^2, 
\nn\\
e^{2\rho_E}&=&\frac{2}{\Lambda}\frac{1}{(x^+-x^-)^2},
 \eea
 where $x^+\equiv t+iz$ and $x^-\equiv t-iz$.
 The embedding coordinates $X$ are defined by:
$X_1\equiv z/2+(-1/\Lambda-t^2)/(2z)$, $X_2\equiv \big(i/\sqrt{|\Lambda|}\big)(t/z)$, and $X_3\equiv z/2+(1/\Lambda+t^2)/(2z)$.
\\

The Euclidean bulk propagator $G_\mathrm{E}$ satisfies the equation
\bea
\big(-\nabla^2+m^2\big) G_\mathrm{E}(y, \tilde{y}; \Delta)
=-\frac{2\sqrt{\pi}\Gamma(\Delta+1/2)}{\sqrt{|\det g_{\mu\nu}|}\Gamma(\Delta)}
\delta^{(2)}(y-\tilde y),
\eea 
where $y$ and $\tilde{y}$ are two bulk-points, and the indices are raised or lowered by the metric $g_{\mu\nu}$.
\\

The Euclidean AdS$_2$ reads:
\bea
X_1&=&\frac{i}{\sqrt{|\Lambda|}}\sinh(\tilde{\rho})\cos(\tau),
\nn\\
X_2&=&\frac{i}{\sqrt{|\Lambda|}}\sinh(\tilde{\rho})\sin(\tau),
\nn\\
X_3&=&\frac{i}{\sqrt{|\Lambda|}}\cosh(\tilde{\rho}),
\nn\\
ds_{2e}^2&=&-dX_1^2-dX_2^2+dX_3^2=-\frac{1}{\Lambda}\big(\sinh^2\tilde{\rho}\ d\tau^2+d\tilde{\rho}^2\big).
\nn\\
\eea
Assuming that the Euclidean bulk propagator only depends on $\tilde{\rho}$, which is also the geodesic distance. The equation of the Euclidean bulk propagator becomes:
\bea
\label{eg}
&&m^2G_{\mathrm E}+\frac{\Lambda}{\sinh\tilde{\rho}}\partial_{\tilde{\rho}}\big(\sinh(\tilde{\rho})\partial_{\tilde{\rho}}G_{\mathrm E}\big)
\nn\\
&=&\Lambda\partial_{\tilde{\rho}}^2G_{\mathrm E}+m^2G_{\mathrm E}+\Lambda\frac{\cosh\tilde{\rho}}{\sinh\tilde{\rho}}\partial_{\tilde{\rho}}G_{\mathrm E}
\nn\\
&=&0.
\eea
Hence we rewrite this Euclidean bulk propagator, $G_{\mathrm E}(\tilde{\rho})\equiv\chi^{\frac{\Delta}{2}} f(\chi)$ and $\chi\equiv \exp(-2\tilde{\rho})$,
plug the Euclidean bulk propagator into \eqref{eg} to obtain $2(\chi-1)\chi f^{\prime\prime}+\big((2\Delta+3)\chi+(-2\Delta-1)\big)f^{\prime}+\Delta f=0$,
and then identify the parameters: $\chi=x$, $a=\Delta$, $b=1/2$, and $c=\Delta+1/2$, in the hypergeometric equation
\bea
&&x(1-x)\frac{d^2}{dx^2}{}_2F_1(a,b;c;x)
\nn\\
&&+\big(c-(a+b+1)x\big)\frac{d}{dx}{}_2F_1(a,b;c;x)
\nn\\
&&-ab\ {}_2F_1(a,b;c;x)
\nn\\
&=&0,
\eea
in which the hypergeometric function ${}_2F_1(a,b;c;x)$ is defined by:
\bea
{}_2F_1(a,b;c;x)\equiv\sum_{n=0}^{\infty}\frac{(a)_n(b)_n}{(c)_n}\frac{x^n}{n!}, \qquad |x|<1,
\eea
where
\bea
 (a)_n\equiv\left\{\begin{array}{ll}
                 1, & \mbox{if $n= 0$}, \\  
                 a(a+1)\cdots (a+n-1), & \mbox{if $n>0$}.
                \end{array} \right.
\eea
We then get the solution of the Lorentzian AdS$_2$ bulk propagator in the non-interacting scalar field theory $f(\chi)={}_2F_1\big(\Delta,1/2;\Delta+1/2;z\big)$
or
$G_{\mathrm E}(\tilde{\rho})=\chi^{\frac{\Delta}{2}}{}_2F_1\big(\Delta,1/2;\Delta+1/2;z\big)$.
\\

The connected four-point function of scalar operators ${\cal F}_{\mathrm{c}}$ can be decomposed as the sum of the conformal blocks, 
\bea
&&{\cal F}_c(\tau_1, \tau_2, \tau_3, \tau_4)
\nn\\
&=&G(\tau_1,\tau_2)G(\tau_3, \tau_4)\sum_{n=1}^{\infty} c_n^2r^{\Delta_n}{}_2F_1(\Delta_n, \Delta_n; 2\Delta_n, r).
\nn\\
\eea
 The cross-ratio $r$ between four points $\tau_1$, $\tau_2$, $\tau_3$, and $\tau_4$ is defined by $r\equiv(\tau_1-\tau_2)(\tau_3-\tau_4)/\big((\tau_1-\tau_3)(\tau_2-\tau_4)\big)$. The quantity $r^{\Delta_n}{}_2F_1(\Delta_n, \Delta_n; 2\Delta_n, r)$ is the conformal block of the connected four-point function associated with an intermediate operator labeled by the index $n$ with the conformal dimension $\Delta_n$, and $c_n^2$ is a non-negative number for each index $n$. The conformal block is a solution to the differential equation following from the requirement that it is annihilated by the Casimir operator \cite{Maldacena:2016hyu} (see also \cite{Dolan:2003hv, SimmonsDuffin:2012uy}). The same equation can be identified with the Klein-Gordon equation in the kinematic space of dS$_2$ (as in \cite{Maldacena:2016hyu}, which is our AdS$_2$ after a sign flip).  
 \\

 We use the following identity of the hypergeometric function
${}_2F_1(a, b; 2b; x)=(1-x)^{-a/2}\cdot{}_2F_1\big(a/2, b-a/2; b+1/2;x^2/(4x-4)\big)$
to obtain the conformal block of the connected four-point function
$r^{\Delta}\cdot{}_2F_1(\Delta, \Delta; 2\Delta; r)=r^{\Delta}(1-r)^{-\Delta/2}\cdot{}_2F_1\big(\Delta/2, \Delta/2; \Delta+1/2;r^2/(4r-4)\big)$.
The convenient identity
${}_2F_1(a, b; 1+a-b;x)=(1-x)^{-a}\cdot{}_2F_1\big(a/2,(1+a)/2-b;1+a-b;-4x/(1-x)^2\big)$
can be used to simplify the Euclidean AdS$_2$ bulk propagator $G_\mathrm{E}(y, \tilde{y}; \Delta)=\chi_{\mathrm{E}}^{\Delta/2}(1-\chi_{\mathrm{E}})^{-\Delta}\cdot{}_2F_1\bigg(\Delta/2,\Delta/2;\Delta+1/2;-4\chi_{\mathrm{E}}/(1-\chi_{\mathrm{E}})^2\bigg)$.
\\

We know $\chi_{\mathrm{E}}=\Lambda^2\xi_{\mathrm{E}}^2\big/\big(1+\sqrt{1-\Lambda^2\xi_{\mathrm{E}}^2}\big)^2$ and $\xi_{\mathrm{E}}=2z\tilde{z}/\big(z^2+\tilde{z}^2+(t-\tilde{t})^2\big)$.
If we consider the Lorentzian AdS$_2$ metric, we need to use the below variables: $\chi_{\mathrm{L}}=\xi_L^2\big((1+\sqrt{1-\xi_L^2})^2\big)$ and $\xi_{\mathrm{L}}=2z\tilde{z}/\big(z^2+\tilde{z}^2-(t-\tilde{t})^2\big)$.
Then we choose $\tau_1=0$, $\tau_2=r$, $\tau_3=1$, and $\tau_4\rightarrow\infty$,
in which we used \eqref{bulkcoord} for the position of the bulk point ($\tau_{3,4}$
for the tilde coordinates). We then have $\xi_{\mathrm{L}}=r/(r-2)$.
Therefore, we obtain
$-4\chi_{\mathrm{L}}/(1-\chi_{\mathrm{L}})^2
=r^2\big/\big(4(r-1)\big)$.
 Hence we show that the Lorentzian AdS$_2$ bulk propagator $ G_\mathrm{L}$ with the conformal dimension $\Delta$ can be reproduced from the conformal block with the conformal dimension $\Delta$, $G_\mathrm{L}(y, \tilde{y}; \Delta)=r^h{}_2F_1(\Delta, \Delta; 2\Delta; r)$.
 \\

\section{Modular Hamiltonian}
\label{sec:3}
The stress tensor also appears in the OPE and the term modular Hamiltonian continues to refer to the corresponding OPE block. In higher dimensions, it has another representation as an integral in a spatial subregion on a time slice and its exponential form gives the reduced density matrix (of the subregion). Consequently, this operator holographically corresponds to the linearized perturbation to the codimension-two minimum surface on a given time slice \cite{Czech:2016xec}. None of these physical features, however, seems to carry over to the current AdS$_2$/CFT$_1$ case. We shall nevertheless see that the variation of OPE block of a stress tensor is supposed to probe the variation of the bulk AdS$_2$ metric. The stress tensor in the CFT$_1$ language is given by the Virasoro generator $L_{-2}$, which is defined from 
\bea
T(\tau)\equiv\sum_{n=-\infty}^{\infty}\frac{L_n}{z^{n+2}}
\eea
acting on the vacuum. We can study the OPE block of the stress tensor since the correspondence established earlier is purely kinematic in CFT$_1$. 
\\
 
 When we consider CFT$_1$, the OPE block \cite{Czech:2016xec} satisfies $z^2(-\partial_z^2+\partial_t^2)B_k(\tau_1, \tau_2)=-\Delta_k(\Delta_k-1) B_k(\tau_1, \tau_2)$,
 where the bulk coordinates, $t$ and $z$, are given by \eqref{bulkcoord}, and $\Delta_k$ is the conformal dimension. One solution for the OPE block of the operator ${\cal O}_k$ is
$ B_k(\tau_1, \tau_2)=\alpha_k\int_{\tau_1}^{\tau_2} dw\ (|w-\tau_2||w-\tau_1|/|\tau_1-\tau_2|)^{\Delta_k-1} {\cal O}_k(w)$, where $\alpha_k$ is a constant for each index $k$. The OPE block of the stress tensor is $ B_T(\tau_1, \tau_2)=(12/c)\cdot\int_{\tau_1}^{\tau_2}dw\ \big((\tau_2-w)(w-\tau_1)/(\tau_2-\tau_1)\big) T(w)$,
 where $c$ is the central charge, and $T(w)$ is the stress tensor. The conformal dimension of stress tensor is two. The variation of boundary stress tensor is given by the Schwarzian derivative:
 \bea
 \delta_f T(w)= \frac{c}{12}\zt{Sch}(f, w)= \frac{c}{12}\lsb\frac {f^{\prime\prime\prime}(\omega)}{f^{\prime}(\omega)} - \frac{3}{2} \bigg(\frac {f^{\prime\prime}(\omega)}{f^{\prime}(\omega)}\bigg)^2\rsb.
\nn\\
 \eea
 Hence the expectation value of the modular Hamiltonian after the infinitesimal transformation $f(\tau) = \tau + \xe(\tau)$ $\big($hence $\xd_f T(\tau)=(c/12)\cdot\epsilon^{\prime\prime\prime}(\tau)\big)$ is $\delta_fB_T(\tau_1, \tau_2)
 =\bigg(\epsilon^{\prime}(\tau_2)+\epsilon^{\prime}(\tau_1)-2\big(\epsilon(\tau_2)-\epsilon(\tau_1)\big)/(\tau_2-\tau_1)\bigg)
 +\cdots$.
 \\

 The two-point function of some scalar operators with the conformal dimension $\Delta=1$ in the CFT$_1$ is given by $G(\tau_1, \tau_2)=\tilde{b}/(\tau_2-\tau_1)^2$,
where $\tilde{b}$ is a constant, when $\tau_2>\tau_1$. The variation of the two-point function from the infinitesimal transformation is:
 \bea
 &&\delta_{f} G(\tau_1, \tau_2)
\nn\\
&\equiv&\tilde{b}\frac{f^{\prime}(\tau_1)f^{\prime}(\tau_2)}{\big(f(\tau_2)-f(\tau_1)\big)^2}
 -\frac{\tilde{b}}{(\tau_2-\tau_1)^{2}}
 \nn\\
 &=&\frac{\tilde{b}}{(\tau_2-\tau_1)^{2}}\bigg(\epsilon^{\prime}(\tau_2)+\epsilon^{\prime}(\tau_1)-2\frac{\epsilon(\tau_2)-\epsilon(\tau_1)}{\tau_2-\tau_1}\bigg)+\cdots
 \nn\\
 &\equiv&\delta_{\epsilon}G(\tau_1, \tau_2)+\cdots.
 \eea
 The two-point function after the transformation is 
\bea
G_f(\tau_1, \tau_2)\equiv\tilde{b}\frac{f^{\prime}(\tau_1)f^{\prime}(\tau_2)}
{\big(f(\tau_2)-f(\tau_1)\big)^2}.
\eea
 Then the modular Hamiltonian is related to the variation of $\delta_{\epsilon}\ln G$: 
 \bea
 \delta_fB_T(\tau_1, \tau_2)&=&\frac{12}{c} \frac{\delta_{\epsilon}G(\tau_1, \tau_2)}{G(\tau_1, \tau_2)}+\cdots
\nn\\
&=&\frac{24}{c}\delta_{\epsilon}\rho_L(\tau_1, \tau_2)+\cdots,
 \eea
 in which we used $\exp(2\rho_L)\equiv 1/(2\Lambda z^2)$ in the final equality. Now we show the following correspondence 
\bea
\frac{f^{\prime}(\tau_1)f^{\prime}(\tau_2)}{\big(f(\tau_2)-f(\tau_1)\big)^2}\longleftrightarrow e^{2\rho_{Lf} (\tau_1,\tau_2)}
\eea 
with $\exp\big(2\rho_{Lf} (\tau_1,\tau_2)\big)$ being the component of the metric 
\bea
ds_{2}^2 = -2\exp\big(2 \rho_{Lf} (v^+,v^-)\big) dv^+ dv^-,
\eea
 where $v^+\equiv t+z\equiv\tau_2$ and $v^-\equiv t-z\equiv\tau_1$.
When the scalar curvature is a negative constant, one should find $\partial_{\tau_1}\partial_{\tau_2}\rho_{Lf}(\tau_1, \tau_2)
\propto \exp\big(2\rho_{Lf} (\tau_1,\tau_2)\big)$.
With the replacement of the metric $\exp\big(2\rho_{Lf}(\tau_1, \tau_2)\big)$ by $f^{\prime}(\tau_1)f^{\prime}(\tau_2)\big/\big(f(\tau_2)-f(\tau_1)\big)^2$,
one can also see that the equation holds
$
(1/2)\cdot\partial_{\tau_1}\partial_{\tau_2}\ln \bigg(f^{\prime}(\tau_1)f^{\prime}(\tau_2)\big/\big(f(\tau_2)-f(\tau_1)\big)^2\bigg)\propto   f^{\prime}(\tau_1)f^{\prime}(\tau_2)\big/\big(f(\tau_2)-f(\tau_1)\big)^2.
$
This observation suggests that the variation $\xd\rho_{L} (\tau_1,\tau_2)$ as a bulk operator corresponds to the modular Hamiltonian. 
\\

To get a more complete picture, we turn to the finite reparameterization $\tau \to f(\tau)$, whose extension into the bulk leads to the following diffeomorphism:
\bea
\label{diffcoord} 
\tilde{v}^+ = f(v^+),\qquad \tilde{v}^- = f(v^-),
\eea 
and the metric component $\exp(2\rho_{Lf}(\tau_1, \tau_2))$ is given by $G_f(\tau_1 -\tau_2)$. Unfortunately, the OPE block only describes non-interacting fields and never senses the change in a background. For instance, the scalar OPE block constructed earlier does not satisfy the Klein-Gordon equation in the new background.
\\

To introduce the coupling with metric, it is necessary to add the OPE blocks of multi-trace operators \cite{Czech:2016xec} involving the stress tensor. In the AdS$_3$/CFT$_2$ case, it has been proposed that all these OPE blocks can be packed into the Virasoro OPE block \cite{Guica:2016pid, Fitzpatrick:2016mtp}. The Virasoro OPE block is constructed by grouping together the contributions to the OPE that are closed under the Virasoro algebra (\ie terms on the right hand side of \eqref{OPEblockdef} that form an irreducible representation). The vacuum Virasoro OPE block in CFT$_2$ corresponds to $\exp(-\xD S)$ with $S$ being the entanglement entropy (also the geodesics length). We expect similar story in CFT$_1$ but the vacuum Virasoro OPE block $\CV(\tau_1, \tau_2; \xD)$ \footnote{Unlike the OPE block the Virasoro OPE block depends on conformal dimensions $\xD_j=\xD_k=\xD$ of the operators $\CO_j,\CO_k$ in the expansion,  and hence we keep the factor $(\tau_1 - \tau_2)^{-2\xD}$.} has now a different bulk interpretation as $\exp\big(2 \xD \rho_{L}(\tau_1, \tau_2)\big)$. The expectation value of $\CV(\tau_1, \tau_2; \xD)$ in the deformed state is proportional to the two-point function $G_f^\xD(\tau_1, \tau_2)$ and agrees with $\exp\big(2 \xD \rho_{Lf}(\tau_1, \tau_2)\big)$. The modular Hamiltonian can be understood as the $\CO(1/c)$ order term in the expansion \cite{Fitzpatrick:2016mtp} 
\bea
&&\CV(\tau_1, \tau_2; \xD)
\nn\\
&=& \frac 1 {(\tau_1-\tau_2)^{2\xD}}\lsb 1+\frac {12\xD}{c} 
\int_{\tau_1}^{\tau_2}dw\ \frac {(\tau_2-w)(w-\tau_1)} {(\tau_2-\tau_1)} T(w)\rsb 
\nn\\
&&+ \CO\lb \frac 1 {c^2}\rb.
\eea
The generic variation on both sides are related as 
\bea
\xd \CV(\tau_1, \tau_2; \xD)=\Delta \frac{\delta_{f} B_T(\tau_1, \tau_2)}{{(\tau_1-\tau_2)^{2\xD}}},
\eea 
which is consistent with the computation of $\delta_{f} B_T(\tau_1, \tau_2)$ above.

\section{Reparametrization}
\label{sec:4}
Now we elaborate on relation between the reparameterization and the deformation of the asymptotic boundary of the two-dimensional spacetime. This appeared in the derivation of the Schwarzian theory \cite{Jensen:2016pah}. More precisely, we consider new coordinates $\tilde{v}^+$ and $\tilde{v}^-$ related to the origin coordinates $v^+$ and $v^-$ by \eqref{diffcoord}. The boundary in the new coordinate lies at:
\bea
\frac {f(v^+)-f(v^-)}{2} = \tilde{z} = \epsilon,\qquad \tilde{t}= \frac {f(v^+)+f(v^-)}{2}. 
\eea
For later convenience, we rewrite $f(x)\equiv x + v(x)$ and the original coordinates $t$ and $z$ can be obtained by solving the following equations in series of $\epsilon$:
\bea
z + \frac {v(t+z)-v(t-z)}{2} = \epsilon, 
\nn\\
\tilde{t}= t + \frac {v(t+z)+v(t-z)}{2}.
\eea
The solution reads:
\bea
\label{perturbsolndiffeo} 
z = \frac {\epsilon} {1+v^{\prime}(t)} + {\cal O}(\epsilon^2),\qquad t + v(t) = \tilde{t} + {\cal O}(\epsilon),
\eea
in which we do the perturbation with respect to $z$.
The second equation to the lowest order is the same as: 
\bea
\tilde{v}^{\pm}=v^{\pm}+v(v^{\pm})=f(v^{\pm}),
\eea
 and hence the coordinate transformation of the light cone coordinates $x^\pm$ is the same as that of the boundary coordinate $t$. Therefore, we can understand the former as the bulk extension of the latter. The diffeomorphism transformation does not change the proper length of the boundary curve at $\tilde{z}=\epsilon$, and therefore the deformation should be the same as in the derivation of the Schwarzian theory from the two-dimensional dilaton gravity theory. We can confirm this easily from the observation that:
\bea
z = \frac{\epsilon} {1+v^{\prime}(t)}+ {\cal O}(\epsilon^2) = \epsilon \frac {d t}{d \tilde t} + {\cal O}(\epsilon^2)\,,     
\eea
in which the second equality is obtained by taking $\tilde t$ derivative of both sides of the second equation in \eqref{perturbsolndiffeo}. This is exactly what we knew in the induced metric $g_{\tilde t\tilde t} = 1/\epsilon^2$, which appeared in the derivation of Schwarzian theory at the given order in $\xe$.

\section{AdS$_2$ Riemann Curvature Tensor}
\label{sec:5}
Based on the above consistent study, we expect that the modular Berry transport can directly probe the AdS$_2$ Riemann curvature tensor \cite{Czech:2017zfq, Czech:2019vih}. Since the modular Hamiltonian is hermitian, this can be diagonalized as $H_{\mathrm{mod}}\equiv U^{\dagger}{\cal D}U$,
where $U$ is unitary, and ${\cal D}$ is a diagonal matrix. The modular Berry transport is $\partial H_{\mathrm{mod}}/\partial\lambda=U^{\dagger}(\partial{\cal D}/\partial\lambda)U +\lbrack (\partial U^{\dagger}/\partial \lambda)U, H_{\mathrm{mod}}\rbrack$, $P_0\big((\partial U^{\dagger}/\partial \lambda)U\big)=0$, where the projection $P_0$ is onto the zero-modes \ie Hermitian operators that commute with $H_{\mathrm{mod}}$. The second equation says that the transport is parallel when the tangent vector is along the horizontal subspace. The vertical subspace is given by the orbit of the gauge group, which in this bundle, is generated by the zero modes that keep the base space, \ie $H_{\mathrm{mod}}$ invariant.
\\

The modular Hamiltonian in CFT$_1$ can be expressed in terms of the SL(2) generators $H_{\mathrm{mod}}=s_1L_1+s_0L_0+s_{-1}L_{-1}$, where $L_{-1}\equiv i\partial_{\tau}$, $L_0\equiv-\tau\partial_{\tau}$, and $L_1\equiv -i\tau^2\partial_{\tau}$, and $s_1\equiv 2\pi/( \tau_2-\tau_1)$, $s_0\equiv -2\pi i(\tau_1+\tau_2)/(\tau_2-\tau_1)$, and $s_{-1}\equiv -2\pi \tau_1\tau_2/(\tau_2-\tau_1)$.
\\

Since one modular Hamiltonian can be mapped to other modular Hamiltonian from the conformal transformation, the equation can reduce to
\bea
\frac{\partial H_{\mathrm{mod}}}{\partial\lambda}
=\bigg\lbrack \frac{\partial U^{\dagger}}{\partial \lambda}U, H_{\mathrm{mod}}\bigg\rbrack, \qquad
P_0\bigg(\frac{\partial U^{\dagger}}{\partial \lambda}U\bigg)=0
\eea
(In our case $U^{\dagger}(\partial{\cal D}/\partial\lambda)U = 0$ as all modular Hamiltonians have the same eigenvalues.).
With the help of the following algebra: 
\bea
\lbrack H_{\mathrm{mod}}, H_{\mathrm{mod}}\rbrack&=&0, 
\nn\\
\lbrack H_{\mathrm{mod}}, \partial_{\tau_1}H_{\mathrm{mod}}\rbrack&=&-2\pi i\partial_{\tau_1}H_{\mathrm{mod}}, 
\nn\\
\lbrack H_{\mathrm{mod}}, \partial_{\tau_2}H_{\mathrm{mod}}\rbrack&=&2\pi i\partial_{\tau_2}H_{\mathrm{mod}},
\eea
we can solve the modular Berry curvature equation, and this leads to
$\partial_{\lambda}H_{\mathrm{mod}}=\lbrack V_{\delta\lambda}, H_{\mathrm{mod}}\rbrack$, where 
\bea
V_{\delta\lambda}\equiv\frac{1}{2\pi i}\big((\partial_{\lambda}\tau_1)(\partial_{\tau_1}H_{\mathrm{mod}}-(\partial_{\lambda}\tau_2)(\partial_{\tau_2}H_{\mathrm{mod}})\big).
\eea
Therefore, we define the covariant derivative $D_{\lambda}H\equiv\partial_{\lambda}H-\lbrack V_{\delta\lambda}, H\rbrack$ and the commutator for derivatives along directions $\lambda=\tau_1$ and $\lambda=\tau_2$ reads 
\bea
\lbrack D_{\tau_1}, D_{\tau_2}\rbrack H=\frac{i}{\pi(\tau_2-\tau_1)^2}\lbrack H_{\mathrm{mod}}, H\rbrack,
\eea
 which leads to the Berry curvature tensor ${\cal R}_{\tau_1\tau_2}\equiv \bigg(i/\big(\pi(\tau_2-\tau_1)^2\big)\bigg)H_{\mathrm{mod}}$.
This also provides the following curvature ${\cal R}_{z\tau}=-\big(i/(2\pi z_0^2)\big)H_{\mathrm{mod}}$. Here we define $\tau_0\equiv(\tau_1+\tau_2)/2$ and $z_0\equiv(\tau_2-\tau_1)/2$. The subscript 0 of $\tau_0$ and $z_0$ means that we fix the variables.
\\

Now we extend the SL(2) generators from the boundary to the bulk for getting the AdS$_2$ Riemann curvature tensor: $L_1=-2i\tau z\partial _z-i(\tau^2+z^2)\partial_{\tau}$, $L_0=-z\partial_z-\tau\partial_{\tau}$, and $L_{-1}=i\partial_{\tau}$. We can see that $H_{\mathrm{mod}}$ as an isometry generator in the bulk keeps the bulk point $(\tau_1, \tau_2)$ fixed and therefore it is essentially the local Lorentz rotation, which is precisely the generator appearing in the bulk spin connection. From the following commutator relations: $\lbrack L_1, \partial_z\rbrack=2i\tau\partial_z+2iz\partial_{\tau}$, $\lbrack L_0, \partial_z\rbrack=\partial_z$, and $\lbrack L_{-1}, \partial_z\rbrack=0$, and $\lbrack L_1, \partial_{\tau}\rbrack=2iz\partial_z+2i\tau\partial_{\tau}$, $\lbrack L_0, \partial_{\tau}\rbrack=\partial_{\tau}$, and $\lbrack L_{-1}, \partial_{\tau}\rbrack=0$, we can find that the diagonal entries of modular Hamiltonian (as a two by two matrix on the tangent vectors) vanish at the point ($\tau_0$, $z_0$), and the off-diagonal ones are symmetric and are $2\pi i$. Therefore, the Riemann curvature at the point $z_0$ (a Lie algebra valued 2-form ${\cal R}_{z\tau} \rmd z \wedge \rmd \tau$ in our convention) is:
\bea
{\cal R}_{z\tau}=-\frac{i}{2\pi z_0^2}H_{\mathrm{mod}}\bigg|_{z=z_0}=\frac{\xS}{z_0^2}=-{\cal R}_{\tau z}.
\eea
where $\xS$ is the Lorentz generator in the vector representation with unit components \ie $\xS^z{}_{\tau} = 1$. The AdS$_2$ Riemann curvature tensor:
\bea
R^{\rho}{}_{\sigma\mu\nu}
&\equiv&\partial_{\mu}\Gamma^{\rho}{}_{\nu\sigma}-\partial_{\nu}\Gamma^{\rho}{}_{\mu\sigma}
+\Gamma^{\rho}{}_{\mu\lambda}\Gamma^{\lambda}{}_{\nu\sigma}-\Gamma^{\rho}{}_{\nu\lambda}\Gamma^{\lambda}{}_{\mu\sigma}, 
\nn\\
\Gamma^{\mu}{}_{\nu\delta}&\equiv&\frac{1}{2}g^{\mu\lambda}(\partial_{\delta}g_{\lambda\nu}+\partial_{\nu}g_{\lambda\delta}-\partial_{\lambda}g_{\nu\delta})
\eea
 exactly corresponds to the curvature ${\cal R}$ at the point $z_0$:
\bea
{\cal R}_{z\tau} \to R^z{}_{\tau z\tau}=\frac{1}{z_0^2}, \qquad {\cal R}_{\tau z}\to R^{z}{}_{\tau\tau z}=-\frac{1}{z_0^2}.
\eea 
Therefore, we use the curvature ${\cal R}$ to probe the AdS$_2$ Riemann curvature tensor.

\section{Outlook}
\label{sec:6}
We related two-boundary points to each bulk point in the Lorentzian\\
 AdS$_2$/CFT$_1$ correspondence. The correspondence provided the most natural counterpart to the codimension-two surface in higher dimensions \cite{Czech:2016xec, deBoer:2016pqk}. In the CFT$_1$ case, the kinematic space directly corresponds to the AdS$_2$ space. This implies that the OPE block is a bulk local operator. Therefore, we can treat the OPE block more easily than in the higher dimensions. Because it is not trivial for the codimension-two surface being a point, we checked the holographic set-up from the conformal block, the modular Hamiltonian, and the reparametrization. The consistent check lets us use the set-up to probe the AdS$_2$ gravity theory conveniently without the extra mapping between the kinematic space and AdS$_2$ spacetime. Finally, as an application of Lorentzian AdS$_2$/CFT$_1$ correspondence, we probed the AdS$_2$ Riemann curvature tensor using the holonomy of the modular Hamiltonian. Because this tensor only has one physical degree of freedom, and we can directly study the AdS$_2$ space, we explicitly confirmed the relation between the modular Berry transport and the curvature. Based on our study, further development from our set-up should be interesting for checking other holographic proposals.

\section*{Acknowledgments}
We would like to thank Bartlomiej Czech for his useful discussion.
Xing Huang acknowledges the support of NWU Starting Grant No.0115/338050048 and the Double First-class University Construction Project of Northwest University. Chen-Te Ma was supported by the Post-Doctoral International Exchange Program and China Postdoctoral Science Foundation, Postdoctoral General Funding: Second Class (Grant No. 2019M652926), and would like to thank Nan-Peng Ma for his encouragement. We would like to thank the National Tsing Hua University, Tohoku University, Okinawa Institute of Science and Technology Graduate University, Yukawa Institute for Theoretical Physics at the Kyoto University, Istituto Nazionale Di Fisica Nucleare - Sezione di Napoli at the Università degli Studi di Napoli Federico II, Kadanoff Center for Theoretical Physics at the University of Chicago, Stanford Institute for Theoretical Physics at the Stanford University, Kavli Institute for Theoretical Physics at the University of California Santa Barbara, Israel Institute for Advanced Studies at the Hebrew University of Jerusalem, Jinan University, and Institute of Physics at the University of Amsterdam.
Discussions during the workshops, ``Novel Quantum States in Condensed Matter 2017'', ``The NCTS workshop on correlated quantum many-body systems: from topology to quantum criticality'', ``String-Math 2018'', ``Strings 2018'', ``New Frontiers in String Theory'', ``Strings and Fields 2018'', ``Order from Chaos'', ``NCTS Annual Theory Meeting 2018: Particles, Cosmology and Strings'', ``The 36th Jerusalem Winter School in Theoretical Physics - Recent Progress in Quantum Field / String Theory'', ``Jinan University Gravitational Frontier Seminar'', ``Quantum Information and String Theory'', ``Strings 2019'', and ``Amsterdam Summer Workshop on String Theory'', were useful to complete this work.


\end{document}